\documentclass[a4paper,review,11pt,nonatbib]{elsarticle}
\journal{International Journal of Forecasting}

\usepackage[a4paper,text={16cm,25cm}]{geometry}
\usepackage[labelsep=period]{caption}

\usepackage{array}
\usepackage{booktabs}
\usepackage{amsthm}
\usepackage{amsmath}
\usepackage{amssymb}
\usepackage{amstext}
\usepackage{amsfonts}
\usepackage{graphicx}
\usepackage{bbm}
\usepackage{bm}
\usepackage{todonotes}

\usepackage{algorithm}
\usepackage{algorithmicx}
\usepackage{algpseudocode}

\usepackage{xurl}
\usepackage{hyperref}
\hypersetup{
  colorlinks=true,
  linkcolor=blue,
  citecolor=blue,
  urlcolor=blue}

\makeatletter
\let\c@author\relax
\makeatother
\usepackage[style=authoryear-comp, backend=biber, uniquename=false, maxnames=9, maxcitenames=2, maxbibnames=99, natbib=true, bibencoding=utf8, dashed=false,terseinits=true, giveninits=true, uniquelist=false, doi=false, isbn=false, url=true, sortcites=false, date=year,safeinputenc]{biblatex}

\DeclareFieldFormat{url}{\texttt{\url{#1}}}
\DeclareFieldFormat[article]{pages}{#1}
\DeclareFieldFormat[inproceedings]{pages}{\lowercase{pp.}#1}
\DeclareFieldFormat[incollection]{pages}{\lowercase{pp.}#1}
\DeclareFieldFormat[article]{volume}{\mkbibbold{#1}}
\DeclareFieldFormat[article]{number}{\mkbibparens{#1}}
\DeclareFieldFormat[article]{title}{\MakeCapital{#1}}
\DeclareFieldFormat[inproceedings]{title}{#1}
\DeclareFieldFormat{shorthandwidth}{#1}
\usepackage{xpatch}
\xpatchbibmacro{volume+number+eid}{\setunit*{\adddot}}{}{}{}
\renewbibmacro{in:}{%
  \ifentrytype{article}{}{%
  \printtext{\bibstring{in}\intitlepunct}}}
\makeatletter
\DeclareDelimFormat[cbx@textcite]{nameyeardelim}{\addspace}
\makeatother

\usepackage{enumitem}
\newlist{steps}{enumerate}{1}
\setlist[steps, 1]{leftmargin = 2cm, label = Step \arabic*:}

\newcommand{\pkg}[1]{{\normalfont\fontseries{b}\selectfont #1}}
\let\proglang=\textsf
\let\code=\texttt
\let\method=\textsf

\newcommand{\specialcell}[2][c]{
  \begin{tabular}[#1]{@{}l@{}}#2\end{tabular}} 

\usepackage[T1]{fontenc}
\usepackage{titlesec}
\titleformat{\paragraph}[runin]{
  \normalfont\bfseries}{\theparagraph}{1em}{}

\usepackage{todonotes}
\usepackage{xcolor}

\begin{document}

\begin{frontmatter}

\title{\bf
Another look at forecast trimming for combinations:\\ robustness, accuracy and diversity
}

\author[mainaddress]{Xiaoqian Wang}
\ead{xiaoqianwang@buaa.edu.cn}

\author[mainaddress]{Yanfei Kang}
\ead{yanfeikang@buaa.edu.cn}

\author[secondaryaddress]{Feng Li\corref{correspondingauthor}}
\cortext[correspondingauthor]{Corresponding author}
\ead{feng.li@cufe.edu.cn}

\address[mainaddress]{School of Economics and Management, Beihang University, Beijing,
  100191, China}
\address[secondaryaddress]{School of Statistics and Mathematics, Central University of Finance
  and Economics, Beijing 102206, China}

\begin{abstract}
  Forecast combination is widely recognized as a preferred strategy over forecast selection due to its ability to mitigate the uncertainty associated with identifying a single ``best'' forecast. Nonetheless, sophisticated combinations are often empirically dominated by simple averaging, which is commonly attributed to the weight estimation error. The issue becomes more problematic when dealing with a forecast pool containing a large number of individual forecasts. In this paper, we propose a new forecast trimming algorithm to identify an optimal subset from the original forecast pool for forecast combination tasks. In contrast to existing approaches, our proposed algorithm simultaneously takes into account the robustness, accuracy and diversity issues of the forecast pool, rather than isolating each one of these issues. We also develop five forecast trimming algorithms as benchmarks, including one trimming-free algorithm and several trimming algorithms that isolate each one of the three key issues. Experimental results show that our algorithm achieves superior forecasting performance in general in terms of both point forecasts and prediction intervals. Nevertheless, we argue that diversity does not always have to be addressed in forecast trimming. Based on the results, we offer some practical guidelines on the selection of forecast trimming algorithms for a target series.
\end{abstract}


\end{frontmatter}

\newpage
\section{Introduction}
\label{sec:intro}

The field of forecasting has seen a great proliferation of literature in both theory and practice \citep{De_Gooijer2006-eg,Petropoulos2022-ft}. Accordingly, the toolbox of forecasting methods has grown in size and sophistication, making two typical ideas, forecast selection and forecast combinations, prevail among forecasters.

Forecast selection aims to identify a single ``best'' forecast for a target time series based on information criteria \citep[e.g.,][]{Kolassa2011-ai}, past forecasting performance \citep[e.g.,][]{Inoue2006-se}, and representativeness of the out-of-sample forecasts \citep{Petropoulos2022-ir}. However, choosing a single forecast out of a set of available forecasts may be misleading because of three sources of uncertainty, namely data uncertainty, model uncertainty, and parameter uncertainty \citep{Petropoulos2018-fw}. To overcome this, an alternative strategy, forecast combinations, moves attention towards finding the optimal weights of combining different forecasts, thus enhancing the forecasting performance through the integration of information gleaned from different sources. A recent review of the extensive literature on forecast combinations since the seminal work of \citet{Bates1969-yj} was provided by \citet{Wang2022-fc}.

Though it is widely established in the forecasting community that combining forecasts is beneficial, the gains from forecast combinations highly depend on several factors, including the quality of the pool of forecasts to be combined and the estimation of combination weights \citep{Timmermann2006-en,Wang2022-fc}. Naturally one would prefer to combine individual forecasts with high accuracy \citep{Mannes2014-dl,Kourentzes2019-na} and sufficient diversity \citep{Batchelor1995-ps,Thomson2019-al,Kang2022-ol} to amplify the benefits of combinations. Alternatively, combination schemes have evolved from simple averaging without weight estimation to sophisticated methods tailoring weights for different individual forecasts. Nonetheless, empirical applications show that simple averaging often dominates sophisticated methods that should (asymptotically) be superior, which is commonly referred to as the ``forecast combination puzzle'' \citep{Stock2004-rq}. The phenomenon is commonly attributed to the weight estimation error \citep{Smith2009-wd,Claeskens2016-pv}, which is so large that it overwhelms the gains from the combination; see \citet{Wang2022-fc} for a recent review of the related literature. The issue is even more problematic when we estimate combination weights for a given pool involving a larger number of individual forecasts.

Instead of combining the complete set of forecasts under consideration, it is intuitively reasonable and widely suggested to combine only a subset of individual forecasts, since there are increasing weight estimation error and decreasing returns to including additional forecasts \citep{Hibon2005-ok,Aiolfi2006-rh,Timmermann2006-en,Geweke2011-xk}. Moreover, a larger pool of individual forecasts may raise the risk of an outlier forecast creeping into the pool and undermining the utility of forecast combinations \citep{Atiya2020-ge}. The additional step that constructs an optimal subset from the full set of available forecasts is called forecast trimming (also known as forecast pooling, subset selection, and combination selection). The main idea of forecast trimming is that \textit{many could be better than all} \citep{Wang2022-fc}. For instance, \citet{Stock1998-np} confirmed the benefits of forecast trimming in forecast combinations when nonlinear models, which tend to produce large forecast errors, are involved in the model pool to be combined. Through forecast trimming, one can mitigate the risk of identifying a single forecast, enhance the quality of the pool of individual forecasts to be combined, and thus amplify the benefits of forecast combinations with improved computational efficiency.

When determining which forecasts should be combined, it is crucial to look at characteristics of the available individual forecasts, among which robustness, accuracy, and diversity are the most frequently emphasized in the literature \citep[see, e.g.,][]{Budescu2015-tu,Thomson2019-al,Atiya2020-ge,Lichtendahl2020-ut}. However, forecast trimming has received very limited research attention so far, and existing algorithms mainly focus on eliminating worst-performing individual forecasts from the forecast pool based on information criteria and some measures of forecast accuracy \citep[e.g.,][]{Granger2004-sw,Mannes2014-dl,Kourentzes2019-na}. Even fewer studies have taken diversity into account to select the optimal subset of individual forecasts. The only two studies we know of that explicitly used diversity for forecast trimming are by \citet{Cang2014-tp} and \citet{Lichtendahl2020-ut} (see Section~\ref{sec:background} for more details). Nevertheless, they selected the optimal subset by examining all possible combinations or by looking only at the forecast error correlations for each pair of individual forecasts. Furthermore, to the best of our knowledge, no existing study has addressed accuracy and robustness of the individual forecasts, as well as diversity of the forecast pool for forecast trimming tasks.

We extend this literature by proposing an algorithm for forecast trimming that addresses robustness, accuracy, and diversity simultaneously. In contrast to \citet{Cang2014-tp} and \citet{Lichtendahl2020-ut}, we pursue a correct trade-off between accuracy and diversity using a new criterion inspired by ambiguity decomposition, rather than dealing with the two issues either in isolation or in sequence. This may be particularly attractive when an available forecast pool contains individual forecasts that perform poorly but have the potential to achieve gains by injecting diversity of knowledge and information. Moreover, our proposed algorithm includes an additional step at the beginning to exclude from the forecast pool the individuals that lack robustness. This additional step removes individuals with a higher risk of bad forecasts. One advantage of our algorithm is that it automatically determines the cut-off point at which we stop eliminating individual forecasts from the pool through a level parameter. Additionally, it is simple and generic, and can be considered as an additional step ahead of the research on forecast combinations. It requires only a small amount of additional computational time due to the calculation simplicity of the required criterion. Our experimental results suggest that the optimal subset identified using our forecast trimming algorithm achieves good performance and robustness in general in terms of both point forecasts and prediction intervals.

We also design another five trimming algorithms that serve as benchmarks to facilitate performance comparisons and future algorithmic development, including one trimming-free algorithm and several trimming algorithms that isolate each one of the three key issues (robustness, accuracy, and diversity). The comparison analysis in our study suggests that we do not necessarily have to address diversity when trimming a forecast pool, even though a stream of research in recent years has discussed and emphasized the importance of diversity. Last, this paper also attempts to provide some guidelines for how to select an appropriate algorithm for forecast trimming when given a time series and its forecast pool.

The rest of the paper is organized as follows. Section~\ref{sec:review} elaborates briefly on the three issues mentioned above, followed by the research gap. In Section~\ref{sec:method}, we introduce the proposed algorithm for forecast trimming and present the benchmarks including trimming-free algorithm and algorithms that isolate each one of the three issues. Section~\ref{sec:experiment} shows the setup of the empirical experiments and the results based on the exponential smoothing family. Section~\ref{sec:diffclass} proceeds by investigating the performance of the trimming algorithms using a forecast pool consisting of models across different model families. Finally, Section~\ref{sec:conclusions} concludes the paper and suggests directions for future work.

\section{Background research and gap}
\label{sec:review}

\subsection{Accuracy, robustness and diversity}
\label{sec:background}

\paragraph{Accuracy.}

Forecast combinations base their performance on the mean level of the accuracy of the individual forecasts to be combined. The individual forecasts should not be very poor, and otherwise combining them would not achieve gains in forecasting performance. Forecast accuracy can be assessed using a variety of error measures, including scale-dependent measures (e.g., the mean absolute error, the mean squared error), measures based on percentage errors (e.g., the mean absolute percentage error, the symmetric mean absolute percentage error), relative measures (e.g., the mean absolute scaled error, the relative mean absolute error), etc. Including a poorly performing forecast in a combination is likely to deteriorate the accuracy of the combined forecast. Therefore, it makes intuitive sense to eliminate the worst performers from the forecast pool based on some performance criteria and to combine only the top performers. For instance, \citet{Kourentzes2019-na} proposed a heuristic called ``forecast islands'' to automatically formulate forecast pools and found it beneficial in terms of the accuracy of the combined forecasts. Specifically, given some appropriate performance criterion, they order the individual forecasts from best to worst and then exclude those forecasts that present a sharp drop in performance by detecting outliers using Tukey's fences approach. The heuristic proposed by \citet{Kourentzes2019-na} is identical to using top $q$ quantiles for forecast combinations, except that the cut-off point of how many quantiles to use is determined automatically rather than arbitrarily.

Discarding a set of worst performers is also the most common strategy in the literature on the ``wisdom of crowds''; this is usually referred to as ``select-crowd'' strategy \citep[see, e.g.,][]{Mannes2014-dl,Goldstein2014-ea,Budescu2015-tu}. Specifically, they seek to improve the quality of the combination of the crowd's estimates by identifying a particular number of best-performing individuals based on the training set. There is diminishing improvement in performance as additional an individuals is included for combination purpose \citep{Armstrong2001-sj,Mannes2014-dl}. The main reason behind the decreasing returns is the increased weight estimation error which overwhelms the marginal gains \citep{Timmermann2006-en}. Combining five (or near to five) top-performing individuals has been shown to achieve good forecasting performance across settings \citep{Makridakis1983-hg,Hora2004-fz,Mannes2014-dl}. Drawing on insights from the wisdom-of-crowds literature, \citet{Goldstein2014-ea} empirically found that smaller, smarter crowds have the potential to beat the wisdom of the whole crowd, suggesting that a potential avenue for future research lies in identifying and tapping into the wisdom of smart sub-crowds.

\paragraph{Robustness.}

\citet{Lichtendahl2020-ut} highlighted the importance of robustness in dealing with forecast combination problems. Given a single time series, the variance of the accuracy across timestamps indicates how robust an individual forecast is to pattern evolution. This relates to the fact that the characteristics of a time series generally change over time, whereas the pattern detected by a forecasting model in the training set may not hold up well in the validation and test sets. On the other hand, when dealing with a set of time series data, the robustness of a given individual forecast can be assessed using the variance of its accuracy across different series. Thus, it reflects the risk of a specific model giving an extremely poor forecast for a given time series. For these reasons, \citet{Lichtendahl2020-ut} suggested balancing the trade-offs between accuracy and robustness when identifying a subset from the available forecast pool for combination purpose.

\paragraph{Diversity.}

The degree of improvement derived from forecast combinations also relies upon the independent information contained in the component forecasts to be combined \citep{Armstrong2001-sj}, which relates to the diversity of the forecast pool. The very simple intuitive explanation is that there would be no performance improvement if identical individual forecasts are combined. \citet{Lichtendahl2013-ws} discussed the strategy for reporting a forecast in a winner-take-all forecasting competition. The optimal strategy is to exaggerate the forecasters' private information and to down-weight any common information. Ideally, we prefer to combine individual forecasts with negatively-related forecast errors so that they would bracket the realization and cancel out the forecast errors, as suggested by \citet{Bates1969-yj}. Unfortunately, this rarely happens in practice. The major obstacle lies in the fact that the individual forecasts are generated from the similar models that are trained for the same task based on the same training data, and thus they tend to be highly positively correlated in most cases. \citet{Atiya2020-ge} documented that one could amplify diversity by using different models based on different assumptions or variables from different sources which affect the variable of interest in different pathways. The results of M4 competition \citep{Makridakis2020-hu} also emphasized the benefits of combining forecasts generated from statistical and machine learning models.

\citet{Mannes2014-dl} and \citet{Thomson2019-al} recognized diversity as one of the two crucial factors that manipulate the quality of the combined forecasts, the other factor being the level of accuracy. The benefits of diversity are confirmed theoretically by \citet{Atiya2020-ge}, who decomposed the Mean Squared Error (MSE) into a bias term and a variance term and found that the extent of the decrease in variance of the forecast combination becomes larger as the correlation coefficients among the individual forecasts decrease. In the context of forecast combinations, some effort has been directed toward using diversity measures as additional inputs to facilitate forecast combinations. For example, \citet{Lemke2010-wn} included features concerning the diversity of the forecast pool when investigating meta-learning for time series forecasting and showed that the introduced diversity features tend to make a combination more successful. \citet{Kang2022-ol} proposed a diversity-based forecast combination approach using only diversity features in a meta-learning framework. They demonstrated that the approach, without extracting sophisticated time series features, outperforms the Feature-based FORecast Model Averaging (FFORMA) approach proposed by \citet{Montero-Manso2020-tq}, which reported the second-best forecast accuracy in the M4 competition.

It was until recently that diversity of the forecast pool was considered for forecast trimming. \citet{Cang2014-tp} designed an optimal subset selection algorithm using mutual information which measures dependence between individual forecasts. The optimal subset is picked out by trying all possible combinations of the individual forecasts, which suffers from large computational cost. \citet{Lichtendahl2020-ut} screened out individual forecasts with low accuracy and highly correlated errors, respectively. However, considering accuracy and diversity in isolation is questionable, as the inclusion of a poorly performing individual forecast in the pool, though potentially harmful to the pool's mean level of accuracy, may still benefit forecast combinations through injecting diversity of knowledge and information.

\subsection{The research gap}
\label{sec:gap}

Current research on forecast trimming focuses on either forecast accuracy of the individual forecasts or diversity of the forecast pool that survives the accuracy screen. To the best of our knowledge, no existing study has simultaneously addressed robustness, accuracy, and diversity of the forecast pool for forecast trimming tasks. In this study, we therefore propose a new algorithm for forecast trimming, taking into account the robustness and accuracy of the individual forecasts, as well as the degree of diversity of the forecast pool. The forecast trimming algorithm often serves as an additional step ahead of forecast combinations. Therefore, it should, in principle, be simple and generic, yet expressive, and not rely on specific combination methods.

This paper has three aims: (i) propose an algorithm for forecast trimming that addresses robustness, accuracy, and diversity simultaneously and demonstrate empirically its performance; (ii) design several benchmark algorithms for performance comparisons and future algorithmic development; and (iii) offer some practical guidelines for choosing an appropriate forecast trimming algorithm for a target time series.

\section{Forecast trimming}
\label{sec:method}

\subsection{Diversity measures}
\label{sec:div}

Diversity is a fundamental issue of ensemble learning in the literature on machine learning --- the success of ensembles is commonly attributed to the degree of disagreement (or diversity) in the individual learners achieved by performing various forms of data manipulations, with bagging and boosting as representativeness. Despite this, there is no agreed formal definition of diversity, and it remains an open research issue \citep{Kuncheva2003-me}. Diversity measures in ensemble learning can generally be divided into two groups: pairwise measures and non-pairwise measures. The first one comprises approaches that assess the pairwise similarity/dissimilarity between two individual learners and then all the pairwise measurements are averaged to portray ensemble diversity, while the second group seeks to measure the overall diversity of the ensemble directly. For an introduction on some representative measures in the ensemble learning context, see \citet{Zhou2012-cy}.

In the forecasting community, much research has gone into encouraging the right degree of diversity when constructing a forecast pool for forecast combinations, while little attention has been paid to the quantification of diversity among different forecasts. In this study, we use the Mean Squared Error for Coherence \citep[MSEC,][]{Thomson2019-al}, also known as Div in \citet{Kang2022-ol}, to assess the degree of diversity between each pair of individual forecasts. Let $f_{i, h}$ be the $h$-th step forecast of the $i$-th forecaster in a given forecast pool, where $i=1,2,\ldots,M$ and $h=1,2,\ldots,H$. The MSEC between the $i$-th and $j$-th individuals in the forecast pool is defined as
\begin{align*}
  \mathrm{MSEC}_{i, j}=\frac{1}{H} \sum_{h=1}^{H}\left(f_{i, h}-f_{j, h}\right)^{2}.
\end{align*}
A value of zero for this measure indicates that the two individuals ($i$ and $j$) have made identical forecasts, and a larger value indicates a higher degree of diversity.

There are two primary reasons for using MSEC to quantify diversity. First, it is clearly feasible to average all the pairwise MSEC values to characterize the overall diversity of the forecast pool available, which helps to avoid concerning only the diversity between a pair of individual forecasts and ignoring their interaction with the remaining individual forecasts when implementing forecast trimming. Moreover, quantifying diversity with respect to MSEC facilitates the design of a new criterion for forecast trimming that balances diversity with accuracy (see Section~\ref{sec:trade-off} for more details). Specifically, MSEC constitutes one of the decomposed components of the overall MSE of a combined forecast, while the other component sheds light on the mean level of accuracy of the forecast pool.

Additionally, we also investigate some other measures for appraising diversity in a forecasting context and give reasons why we do not consider them in this study. Given two individual forecasts, the diversity is usually formulated in terms of the correlation coefficient between their forecast errors \citep{Lichtendahl2020-ut}. The lower the correlation, the larger the diversity. However, given a forecast pool, averaging the correlation coefficients across pairs of individual forecast errors makes no sense as, mathematically, correlations are not additive \citep{Achen1977-cc} and thus the result can not be utilized to quantify the overall diversity of the pool. For this reason, this pairwise measure can not be directly applied to perform forecast trimming that requires comprehensive consideration of the degree of diversity. The mutual information introduced by \citet{Cang2014-tp} also shares the same problem. Besides, in light of the clustering combination schemes \citep[see, e.g.,][]{Aiolfi2006-rh} that first sort individual forecasts into clusters on the basis of their past forecasting performance, combine forecasts within each cluster, and then estimate the combination weights for these clusters, some different measures have been proposed to quantify diversity, such as the number of individual forecasts in best-performing cluster and the distance of the means of the top two performing clusters \citep{Lemke2010-wn}. These diversity measures, however, depend heavily on the clustering algorithm used, which would dramatically increase algorithm complexity when applied to forecast trimming.

\subsection{Trade-off between accuracy and diversity}
\label{sec:trade-off}

It is desired that the available forecast pool comprises a set of individual forecasts that are both accurate and diverse to achieve effective combinations. Combining some accurate individual forecasts with some relatively poor ones is often a strategy superior to combining only accurate ones due to the complementary benefits of diversity. Ultimately, instead of pursuing accuracy or diversity in isolation, the success of forecast combinations lies in forming a forecast pool exhibiting a correct trade-off between accuracy and diversity. Therefore, it is imminent to propose a new trimming criterion that reflects not only the level of accuracy but also the the degree of diversity of the forecast pool used for combination.

For a given time series $\left\{y_{t}, t=1,2,\ldots,T\right\}$, \citet{Kang2022-ol} showed that the overall MSE of a weighted combined forecast, $\mathrm{MSE}_{comb}$, can be decomposed into component measures that involve accuracy (performance) and diversity (coherence), as follows:
\begin{align}
  \mathrm{MSE}_{comb} &=\sum_{i=1}^{M}w_{i}\left[\frac{1}{H} \sum_{h=1}^{H}\left(f_{i, h}-y_{h}\right)^{2}\right] - \sum_{i=1}^{M-1}\sum_{j=2, j>i}^{M}w_{i}w_{j}\left[\frac{1}{H} \sum_{h=1}^{H}\left(f_{i, h}-f_{j, h}\right)^{2}\right] \nonumber \\
  &= \sum_{i=1}^{M}w_{i}\mathrm{MSE}_{i} - \sum_{i=1}^{M-1}\sum_{j=2, j>i}^{M}w_{i}w_{j}\mathrm{MSEC}_{i, j}, \label{eq:decomp}
\end{align}
where $w_{i}$ is the combination weight (assuming a static weight) assigned to the $i$-th individual forecast in a forecast combination task, and $\mathrm{MSE}_{i}$ is the mean squared error of the $i$-th individual forecast. This decomposition is inspired by ambiguity decomposition \citep{Krogh1994-ne} in the literature on machine learning. It indicates that a link exists between the accuracy measures for individual forecasts and the diversity measures between all pairs of forecasts.

Drawing on insights from the above decomposition, we propose a new criterion for trimming an available forecast pool to identify an optimal subset for the subsequent research on effective forecast combinations, denoted as Accuracy-Diversity Trade-off (ADT). The ADT criterion is given by
\begin{align}
  \mathrm{ADT} \quad &= \quad \mathrm{AvgMSE} \quad\quad - \quad \kappa\mathrm{AvgMSEC} \nonumber \\
  &=\underbrace{\frac{1}{M}\sum_{i=1}^{M}\mathrm{MSE}_{i}}_{\text{mean level of accuracy}} - \quad \kappa\underbrace{\frac{1}{M^{2}}\sum_{i=1}^{M-1}\sum_{j=2, j>i}^{M}\mathrm{MSEC}_{i, j}}_{\text{overall diversity}}, \label{eq:ADT}
\end{align}
where $\kappa$ is a scale factor and $\kappa \in \left[0,1\right]$. Note that if $\kappa>1$ the criterion would be negative with arbitrarily large magnitude. Similar to the decomposition form of $\mathrm{MSE}_{comb}$ in Equation~\eqref{eq:decomp}, the ADT criterion consists of two parts: one refers to the mean level of accuracy (in terms of forecast errors); and the other to the overall diversity of the available forecast pool. However, the ADT criterion introduces the following new considerations.

\begin{enumerate}
  \item The ADT criterion is a simplified loss function with equal weights. In our study, we restrict our focus to equal weights (simple averaging) for two reasons. First, forecast trimming is commonly applied as an additional step ahead of forecast combinations. In this regard, the criterion used for forecast trimming should be simple and generic, and should not rely on specific weight estimation method so that the resulting optimal subset can be used for a variety of subsequent studies on forecast combinations. Second, the ``forecast combination puzzle'' indicates that simple averaging sets a tough benchmark, with few combination schemes outperforming it.
  \item A scale factor, $\kappa$, is introduced in the ADT criterion to explicitly control the emphasis on the diversity component. The higher the $\kappa$ value, the more emphasis is placed on diversity when measuring the quality of a forecast pool. When $\kappa=0$, the quality of a forecast pool is considered to depend only on the average of the individuals' accuracy. When $\kappa=1$, the ADT criterion addresses both accuracy and diversity issues and it reduces to the MSE value of an equally combined forecast. Moreover, given a new series and its forecast pool, the optimal $\kappa$ can be identified either artificially for various applications or automatically based on historical performance.
  \item The idea of the ADT criterion is in line with research on ensemble with Negative Correlation Learning \citep[NCL,][]{Liu1999-nc,Brown2005-ma} in the context of machine learning, where the error function contains two parts: the empirical risk function of the individual network and the correlation penalty function. Through the correlation penalty term, NCL allows all the individual networks in an ensemble to train simultaneously and interactively, thus creating an ensemble with negatively correlated networks. Instead of including a correlation penalty term, the ADT criterion introduces a diversity incentive term to achieve the same goal of making individual forecasts to be combined as diverse as possible. Unlike NCL, the ADT criterion is utilized to trim a given forecast pool, rather than to influence the individual model training process. Thus, the ADT criterion takes the forecast pool as given and is applicable to forecasts of different families, including statistical, machine learning, and judgmental forecasts.
\end{enumerate}

\subsection{The RAD algorithm}
\label{sec:RAD}

We propose a new forecast trimming algorithm, denoted as RAD, for selecting forecasts from a given forecast pool to formulate an optimal subset, addressing Robustness, Accuracy, and Diversity simultaneously. To apply the \method{RAD} algorithm, we first divide the available in-sample data into the training set $D_{train}$ and the validation set $D_{valid}$. The training set is used to fit statistical or machine learning models, even to provide human forecasters with insights into data patterns. The forecasts from these fitted models or forecasters are then evaluated against the validation set, whose length is the same as the required out-of-sample horizon $H$, and then used to identify an optimal subset from the complete set of forecasts under consideration using the \method{RAD} algorithm. Whereas the forecasts produced for the out-of-sample periods are used to validate the optimal subset obtained.

Let $\bm{F} = \left[\bm{f}_{1}, \bm{f}_{2}, \ldots, \bm{f}_{M}\right]^{\prime}$ be an $M \times H$ matrix representing the forecast pool available, where $\bm{f}_{i}=\left(f_{i, 1}, f_{i, 2}, \ldots, f_{i, H}\right)^{\prime}$, $\left(1 \leqslant i \leqslant M\right)$ is the $i$-th individual forecasts on the validation set $D_{valid}$, $H$ is the size of the validation set $D_{valid}$, and $M$ is the total number of the individual forecasts in the original forecast pool. Then the \method{RAD} algorithm is described as below.

\begin{steps}
  \item Set the initial selected individual forecaster set $\mathbb{S} = \left\{1, 2, \ldots, i, \dots, M\right\}$ which is a complete set under consideration.
  \item Apply Tukey's fences approach to exclude from the forecaster set $\mathbb{S}$ the individuals that lack robustness. Specifically, for each individual forecaster $i$ ($1 \leqslant i \leqslant M$) we calculate the variance of absolute errors between the its forecasts $f_{i, h}$ and the actual values $y_{h}$ ($1 \leqslant h \leqslant H$) on the validation set $D_{valid}$, and remove individual forecasters with the variance of absolute errors exceeding $Q_{3}+1.5\left(Q_{3}-Q_{1}\right)$, where $Q_{1}$ and $Q_{3}$ are the first and third quantiles of the respective values across individual forecasters from the set $\mathbb{S}$.
  \item Use Equation~\eqref{eq:ADT} with the scale factor $\kappa$ set to $1$ to calculate the ADT criterion of the forecaster set $\mathbb{S}$ based on the individual forecasts remained in $\mathbb{S}$ and the actual values on the validation set $D_{valid}$, denoted as $\mathrm{ADT}_{0}$.
  \item For each individual forecaster $i$ in the forecaster set $\mathbb{S}$, calculate the ADT value of the remaining set after removing $i$ from $\mathbb{S}$, and find the minimum ADT value $\mathrm{Min}_{i}\mathrm{ADT}(\mathbb{S} \backslash \{i\})$ among all $i$.
  \item Exclude from the forecaster set $\mathbb{S}$ the individual forecasters corresponding to the minimum ADT value $\mathrm{Min}_{i}\mathrm{ADT}(\mathbb{S} \backslash \{i\})$.
  \item Calculate the ADT value for the updated forecaster set $\mathbb{S}$.
  \item Repeat Step 4, Step 5, and Step 6 until there is non-significant reduction of the ADT value for $\mathbb{S}$ compared to the previous one or until $\mathbb{S}$ contains only two forecasters. Thus, the resulting $\mathbb{S}$ is the optimal subset identified from all individual forecasters available, and then their forecasts produced for the test set should be used for the subsequent studies on forecast combinations.
\end{steps}

The \method{RAD} algorithm obviously uses a backward selection strategy in trimming a given forecast pool. A level parameter $\delta$ ($\delta \geqslant 0$) is introduced in \textbf{Step 7} to identify whether the percentage decrease in the ADT value for $\mathbb{S}$ is significant. Specifically, the difference in the ADT values between the forecaster sets $\mathbb{S}$ at consecutive iterations, $k-1$ and $k$, is significant if $\frac{\mathrm{ADT}\left(\mathbb{S}^{k-1}\right) - \mathrm{ADT}\left(\mathbb{S}^{k}\right)}{\mathrm{ADT}\left(\mathbb{S}^{k-1}\right)} \geqslant \delta$. The greater the $\delta$ value, the larger the ADT reduction required to remove an individual forecast from the forecast pool. Different optimal subsets may be obtained if a different $\delta$ level is used in \textbf{Step 7}. We suggest $\delta = 0.05$ as a standard parameter value and a sensitivity analysis on the level parameter is discussed in Section~\ref{sec:analysis}.

\subsection{Benchmark algorithms}
\label{sec:benchmark}

We present five benchmark algorithms for forecast trimming that are implemented in our study; they are described in Table~\ref{tab:algorithms}, together with the \method{RAD} algorithm. They include a trimming-free algorithm, three algorithms that focus on isolating each one of the three key issues (i.e., robustness, accuracy and diversity) involved in selecting a forecast pool for forecast combinations, and an automatic \method{RAD} algorithm which automatically determine the value of the scale factor $\kappa$ based on historical forecasting performance.

\begin{table}[!ht]
  \linespread{0.9}\selectfont
  \centering
  \caption{Summary of forecast trimming algorithms considered along with different issues addressed by each algorithm.}
  \scalebox{0.9}{
  \setlength\extrarowheight{6pt}
  \begin{tabular}{lp{0.5\columnwidth}ccc}
  \toprule
  Algorithm & Description & Robustness & Accuracy & Diversity \\
  \midrule
  \method{None} & Do not trim any individuals from the original forecast pool. & & & \\
  \cmidrule(lr){2-5}
  \method{R} & Exclude only the individuals that lack robustness from the original forecast pool. In other words, only \textbf{Step 2} in the \method{RAD} algorithm is conducted. & $\checkmark$ & & \\
  \cmidrule(lr){2-5}
  \method{A} & Exclude only the individuals with relatively low forecast accuracy from the original forecast pool. Specifically, \textbf{Step 2} in the \method{RAD} algorithm is removed, and $\mathrm{AvgMSE}$, rather than ADT, is used as the criterion in \textbf{Steps 3-7}. & & $\checkmark$ & \\
  \cmidrule(lr){2-5}
  \method{D} & Exclude only the individuals whose departure would result in a significant increase in $\mathrm{AvgMSEC}$ from the original forecast pool. Specifically, \textbf{Step 2} in the \method{RAD} algorithm is removed, and negative $\mathrm{AvgMSEC}$, rather than ADT, is used as the criterion in \textbf{Steps 3-7}. & & & $\checkmark$ \\
  \cmidrule(lr){2-5}
  \method{RAD} & Address robustness, accuracy and diversity simultaneously when implementing forecast trimming, as described in Section~\ref{sec:RAD}. & $\checkmark$  & $\checkmark$ & $\checkmark$ \\
  \cmidrule(lr){2-5}
  \method{AutoRAD} & The only difference from the \method{RAD} algorithm is that the scale factor $\kappa$ is automatically identified as the one that yields an optimal subset with the minimum MSE value of the simple average among all pre-set values of $\kappa$. In this study, we consider $K = \{0, 0.1, 0.2, \ldots, 1\}$ as the set of alternative $\kappa$ values. & $\checkmark$ & $\checkmark$ \\
  \bottomrule
  \end{tabular}}
  \label{tab:algorithms}
\end{table}

\begin{figure}[!ht]
  \centering
  \includegraphics[width=\textwidth]{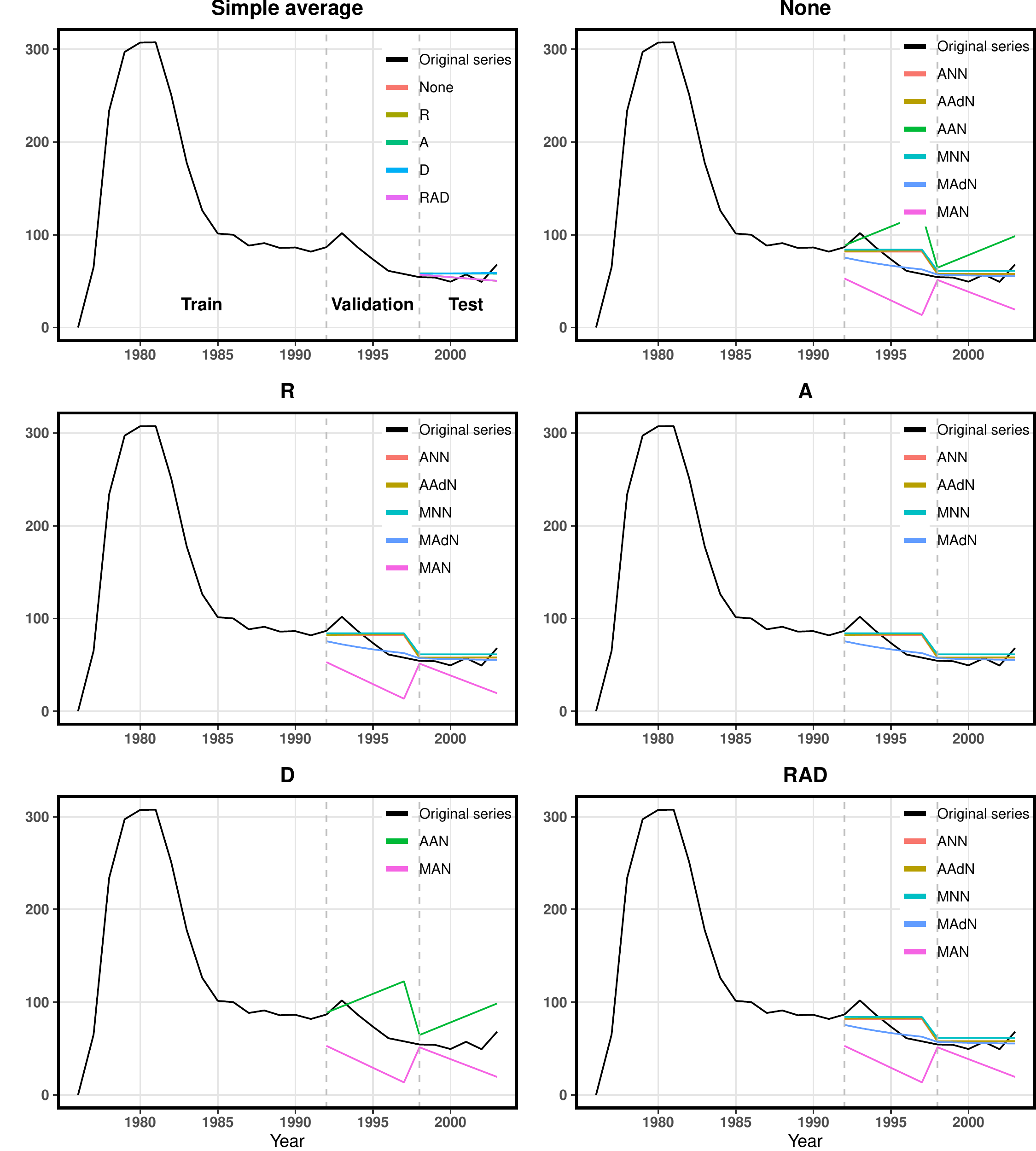}
  \caption{An example showing the individual models retained as well as the simple average forecasts after implementing different forecast trimming algorithms. The vertical dashed lines are used to split the series into the training, validation, and test sets. For this example, we use the $25^{\text{th}}$ yearly series of the M competition data set, Y25, and a forecast pool containing six ETS models.}
  \label{fig:example_series}
\end{figure}

We apply these forecast trimming algorithms to the yearly time series with identification Y25 from the M competition \citep{Makridakis1982-hb}. Six models from the exponential smoothing (ExponenTial Smoothing; ETS) family are included in the original forecast pool, each of which consists of three terms: error, trend, and seasonality. For example, \method{MAdN} refers to an ETS model with multiplicative error, additive damped trend, and no seasonality. Figure~\ref{fig:example_series} shows the optimal subsets selected based on the validation set by using different forecast trimming algorithms (panels other than the top left panel) as well as their simple averages of the individual forecasts on the test set (top left panel). We observe that in this example, \method{R} excludes individual models with larger variance of forecast errors, \method{A} selects individual models with higher accuracy, while \method{D} retains individual models with higher degree of diversity. Additionally, \method{RAD} addresses robustness, accuracy, and diversity simultaneously. As a result of the trade-off between accuracy and diversity, no model is further trimmed after excluding models with larger variance of forecast errors so that the optimal subset obtained by \method{RAD} is identical to the one obtained by implementing \method{R}.

\section{Empirical investigation}
\label{sec:experiment}

\subsection{Data}
\label{sec:data}

To empirically compare the six forecast trimming algorithms presented in Table~\ref{tab:algorithms}, we use the yearly, quarterly, monthly, weekly, daily, and hourly time series from the three most famous forecasting competitions, the M \citep{Makridakis1982-hb}, M3 \citep{Makridakis2000-he}, and M4 \citep{Makridakis2020-hu} competitions. Each series consists of in-sample and out-of-sample (test) sets of observations. The in-sample data is further divided into the training set and the validation set of the same length as the test set (see Section~\ref{sec:RAD} for more details). Time series with training set that comprises fewer than two observations, and time series that are constant over the training set, are eliminated from the data set we consider. In total, we consider 103,826 series of various frequencies and lengths. The forecast horizons of the yearly, quarterly, monthly, weekly, daily, and hourly series are 6, 8, 18, 13, 14, and 48, respectively. The corresponding periodicities (lengths of seasonal cycle) are 1, 4, 12, 52, 7, and 168. These data sets are publicly available in the \proglang{R} packages \pkg{Mcomp} \citep[version 2.8,][]{rMcomp} and \pkg{M4comp2018} \citep[version 0.2.0,][]{rM4comp2018}.

\subsection{Design}
\label{sec:design}

In this section, the forecast pool we consider consists of a set of ETS models. ETS is one of the most popular univariate forecasting methods and is widely used in recent studies on forecast selection \citep{Talagala2018-ms,Meira2021-pi,Petropoulos2022-ir}, forecast trimming \citep{Kourentzes2019-na}, and forecast combinations \citep{Petropoulos2018-fw,Montero-Manso2020-tq,Kang2022-ol,Wang2022-un} due to its robust forecasting performance. We do not involve multiplicative trend models or models with additive error and multiplicative seasonal components in our forecast pool. Therefore, our forecast pool consists of six models for non-seasonal (yearly) series, and 15 models for seasonal series. The \code{ets()} function of the \pkg{forecast} \proglang{R} package \citep[version 8.16,][]{rforecast} is used to model yearly, quarterly, monthly, and daily series, while the \code{es()} function of the \pkg{smooth} \proglang{R} package \citep[version 3.1.5,][]{rsmooth} is used instead to model weekly and hourly series (with lengths of seasonal cycle greater than 24).

For each forecast trimming algorithm, we evaluate the quality of the selected optimal subset using an equal-weighted combination (i.e., a simple average) across all individual forecasts remained in the subset. The equal-weighted combinations are applied to the point forecasts as well as to the prediction intervals obtained from each model in the optimal subset, although the subset is identified using only point forecasts. Here, we consider a simple average instead of an unequal-weighted combination for two primary reasons. First, the choice of weight estimation schemes is subjective. Different criteria are used in different forecast trimming algorithms to recognize an optimal subset from a given forecast pool. It is unfair and improper to compare the performance of these algorithms by combining the individual forecasts with weights that reflect the values of a specific criterion. A second reason is the surprising robustness and superior forecasting performance of simple averaging. Simple averaging sets a rigorous benchmark for the subsequent research on weighted combination methods. The study of weighted combinations based on an optimal subset whose simple averaging performs well is often promising.

Prediction intervals provided by some ETS models may be unrealistically wide. Therefore, we include a pre-step to exclude models that produce outlier prediction intervals from the forecast pool, which is similar to the treatment applied by \citet{Meira2021-pi} and \citet{Petropoulos2022-ir}. Specifically, we eliminate individual models whose lower (or upper) bound of the prediction interval for the furthest horizon is lower than $Q_{1}-1.5\left(Q_{3}-Q_{1}\right)$ (or greater than $Q_{3}+1.5\left(Q_{3}-Q_{1}\right)$), where $Q_{1}$ and $Q_{3}$ are the first and third quantiles of the respective values across models in the given forecast pool. In addition, we exclude models with the lower bound greater than the upper bound for the furthest horizon.

\subsection{Evaluation metrics}
\label{sec:metrics}

To assess the point forecast performance, we consider three measures --- the Mean Absolute Scaled Error \citep[MASE,][]{Hyndman2006-an}, the symmetric Mean Absolute Percentage Error \citep[sMAPE, see for example,][]{Makridakis2000-he}, and the bias measured by the average signed error scaled by the in-sample mean \citep[see for example,][]{Petropoulos2022-ir}. MASE and sMAPE are widely used in many forecasting competitions such as M3 and M4 competitions. These three metrics can be calculated as
\begin{align*}
  \mathrm{MASE} &= \frac{1}{H} \frac{\sum_{t=T+1}^{T+H}\left|y_{t}-f_{t}\right|}{\frac{1}{T-s} \sum_{t=s+1}^{T}\left|y_{t}-y_{t-s}\right|}, \\
  \mathrm{sMAPE} &= \frac{200}{H} \sum_{t=T+1}^{T+H} \frac{\left|y_{t}-f_{t}\right|}{\left|y_{t}\right|+\left|f_{t}\right|}, \\
  \mathrm{Bias} &= \frac{1}{H} \frac{\sum_{t=T+1}^{T+H}\left(y_{t}-f_{t}\right)}{\frac{1}{T} \sum_{t=1}^{T} y_{t}},
\end{align*}
where $y_{t}$ and $f_{t}$ are the actual observation and the forecast at time period $t$, $T$ is the length of the historical observations (sample size), $H$ is the required forecast horizon, and $s$ is the length of seasonal cycle. The values of these three metrics can be averaged across time series because they are scale-independent. Lower values for sMAPE, MASE and absolute bias are better.

The performance of prediction intervals is measured in terms of the Mean Scaled Interval Score \citep[MSIS,][]{Gneiting2007-st}, coverage, upper coverage, and spread. The MSIS value can be calculated as
\begin{align*}
  \mathrm{MSIS} = \frac{1}{H} \frac{\sum_{t=T+1}^{T+H}\left(\left(u_{t}-l_{t}\right)+\frac{2}{\alpha}\left(l_{t}-y_{t}\right) \mathbf{1}\left\{y_{t}<l_{t}\right\}+\frac{2}{\alpha}\left(y_{t}-u_{t}\right) \mathbf{1}\left\{y_{t}>u_{t}\right\}\right)}{\frac{1}{T-s} \sum_{t=s+1}^{T}\left|y_{t}-y_{t-s}\right|},
\end{align*}
where $l_{t}$ and $u_{t}$ are lower and upper bounds of the generated $100\left(1-\alpha\right)\%$ prediction interval at time period $t$, and $\mathbf{1}$ is an indicator function that returns 1 if the condition holds and 0 otherwise. MSIS balances the width of the produced prediction interval and the penalty for the actual value lying outside the interval. Coverage refers to the percentage of times that the actual values lie inside the prediction intervals. Upper coverage, a proxy for achieved service level, refers to the percentage of times that the actual values are not larger than the upper bounds of the prediction intervals. Spread of the prediction intervals, a proxy for holding costs, measures the average intervals scaled similarly to bias. In this paper, we produce prediction intervals at a $95\%$ confidence level ($\alpha=0.05$). Lower values for MSIS and spread are better. The target values for coverage and upper coverage are $95\%$ and $97.5\%$, respectively.

\subsection{Forecast combination results}
\label{sec:results}

Table~\ref{tab:results_mcomp} presents the performance of equal-weighted combinations across individual forecasts from the optimal subsets identified by different forecast trimming algorithms as shown in Table~\ref{tab:algorithms} for each competition data set and measure. Also the bottom panel offers the overall results across all data sets. For each data set and measure, entries in bold highlight the best forecast trimming algorithm.

\begin{table}[!ht]
  \centering
  \caption{The average forecasting performance of each forecast trimming algorithm for each data set and measure. The bottom panel also reports the overall results across all data sets. The best trimming algorithm is boldfaced for each data set and measure.}
  \scalebox{0.8}{
  \setlength\extrarowheight{-2pt}
  \begin{tabular}{ccp{0.125\columnwidth}p{0.125\columnwidth}p{0.125\columnwidth}p{0.125\columnwidth}p{0.125\columnwidth}p{0.125\columnwidth}}
  \toprule
   & & \multicolumn{6}{c}{\textbf{Simple Average}} \\
  \cmidrule(lr){3-8}
  Data set & Measure & \method{None} & \method{R} & \method{A} & \method{D} & \method{RAD} & \method{AutoRAD} \\
  \midrule
  M & MASE & 1.693 & 1.685 & \textbf{1.598} & 1.751 & 1.600 & 1.601 \\
   & sMAPE & 16.157 & 16.062 & \textbf{15.242} & 16.663 & 15.484 & 15.246 \\
   & MSIS & \textbf{18.702} & 18.739 & 19.398 & 19.249 & 19.044 & 19.228 \\
   & Coverage & 0.877 & 0.874 & 0.852 & \textbf{0.879} & 0.858 & 0.854 \\
   & Upper coverage & 0.916 & 0.915 & 0.908 & \textbf{0.917} & 0.911 & 0.909 \\
   & Spread & 0.980 & 0.974 & \textbf{0.875} & 1.004 & 0.889 & \textbf{0.875} \\
   & Bias & 0.071 & 0.071 & \textbf{0.058} & 0.071 & \textbf{0.058} & \textbf{0.058} \\
  \midrule
  M3 & MASE & 1.387 & \textbf{1.383} & 1.401 & 1.443 & 1.399 & 1.399 \\
   & sMAPE & 13.399 & \textbf{13.355} & 13.401 & 13.997 & 13.383 & 13.371 \\
   & MSIS & \textbf{11.424} & 11.444 & 13.373 & 11.682 & 13.103 & 13.181 \\
   & Coverage & 0.928 & 0.927 & 0.905 & \textbf{0.931} & 0.911 & 0.909 \\
   & Upper coverage & 0.948 & 0.948 & 0.939 & \textbf{0.950} & 0.942 & 0.942 \\
   & Spread & 0.844 & 0.838 & \textbf{0.785} & 0.890 & 0.798 & 0.792 \\
   & Bias & 0.014 & 0.013 & \textbf{0.003} & 0.013 & \textbf{0.003} & \textbf{0.003} \\
  \midrule
  M4 & MASE & 1.574 & 1.535 & 1.521 & 1.758 & \textbf{1.520} & \textbf{1.520} \\
   & sMAPE & 12.284 & 12.239 & 12.154 & 12.708 & \textbf{12.148} & 12.149 \\
   & MSIS & 24.729 & 18.005 & 14.300 & 48.813 & \textbf{14.219} & 14.245 \\
   & Coverage & \textbf{0.933} & 0.932 & 0.918 & 0.929 & 0.921 & 0.920 \\
   & Upper coverage & \textbf{0.954} & \textbf{0.954} & 0.951 & 0.950 & 0.952 & 0.952 \\
   & Spread & 1.408 & 1.105 & \textbf{0.892} & 2.461 & 0.904 & 0.898 \\
   & Bias & 0.027 & 0.033 & 0.021 & \textbf{0.010} & 0.022 & 0.022 \\
  \midrule
  Overall & MASE & 1.570 & 1.533 & 1.519 & 1.749 & \textbf{1.518} & \textbf{1.518} \\
   & sMAPE & 12.352 & 12.306 & 12.218 & 12.782 & 12.214 & \textbf{12.212} \\
   & MSIS & 24.308 & 17.834 & 14.324 & 47.516 & \textbf{14.235} & 14.264 \\
   & Coverage & \textbf{0.933} & 0.931 & 0.917 & 0.929 & 0.921 & 0.919 \\
   & Upper coverage & \textbf{0.953} & \textbf{0.953} & 0.950 & 0.950 & 0.952 & 0.951 \\
   & Spread & 1.389 & 1.097 & \textbf{0.889} & 2.404 & 0.901 & 0.895 \\
   & Bias & 0.027 & 0.033 & 0.021 & \textbf{0.011} & 0.022 & 0.022 \\
  \bottomrule
  \end{tabular}}
  \label{tab:results_mcomp}
\end{table}

We observe that, compared with \method{None}, \method{R} leads to improved point forecast accuracy for each data set, which however comes at a price in terms of interval forecast accuracy. In other words, trimming a given forecast pool by only addressing robustness is beneficial for point forecasting but not for interval forecasting. \method{D} consistently results in a very poor performance compared with \method{None} because \method{D} unilaterally pursues pairs of individual forecasts with large differences (diversity), increasing the risk of reserving very poor individual forecasts in the optimal subset.

Overall, \method{RAD} and \method{AutoRAD} outperform the benchmark forecast trimming algorithms for both point forecasts and prediction intervals. Specifically, \method{A}, \method{RAD}, and \method{AutoRAD} offer better results compared to \method{None}, \method{R}, and \method{D} in terms of the overall values of MASE, sMAPE, and MSIS. More importantly, \method{RAD} and \method{AutoRAD} stand out as the top two performing algorithms when we focus on the mean forecast errors across all data sets. This superiority is particularly evident when considering prediction intervals. Moreover, the forecast errors in terms of MASE, sMAPE, and MSIS for \method{RAD} are $3.31\%$, $1.12\%$, and $41.44\%$, respectively, lower than for \method{None}, even though these trimming algorithms only use the simplest equal-weighted combination strategy. \method{None} reports the best coverage and upper coverage, but this performance comes at a price in terms of spread of prediction intervals and bias. Whereas \method{RAD} and \method{AutoRAD} achieve a balance between coverage and interval spread. Overall, trimming algorithms that simultaneously address robustness, accuracy, and diversity provide better forecasting performance when considering equal-weighted combinations compared with the trimming-free algorithm and the trimming algorithms that isolate each one of the three issues.

\begin{figure}[!ht]
  \centering
  \includegraphics[width=\textwidth]{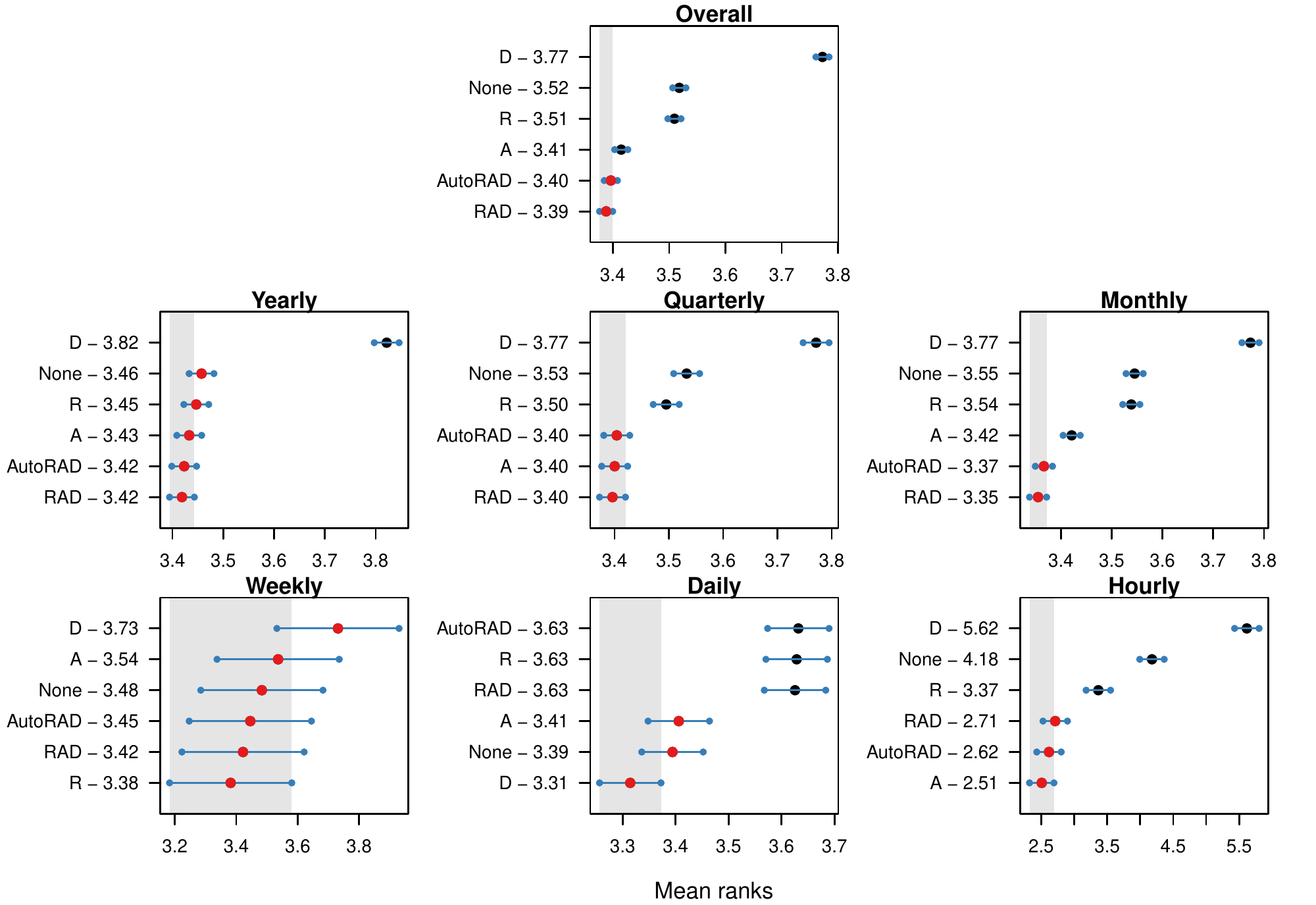}
  \caption{MCB tests on the ranks of the MASE values of the simple average forecasts remained after using \method{None}, \method{R}, \method{A}, \method{D}, \method{RAD} and \method{AutoRAD} for each data frequency separately and across all frequencies (Overall).}
  \label{fig:mcb_mase}
\end{figure}

\begin{figure}[!ht]
  \centering
  \includegraphics[width=\textwidth]{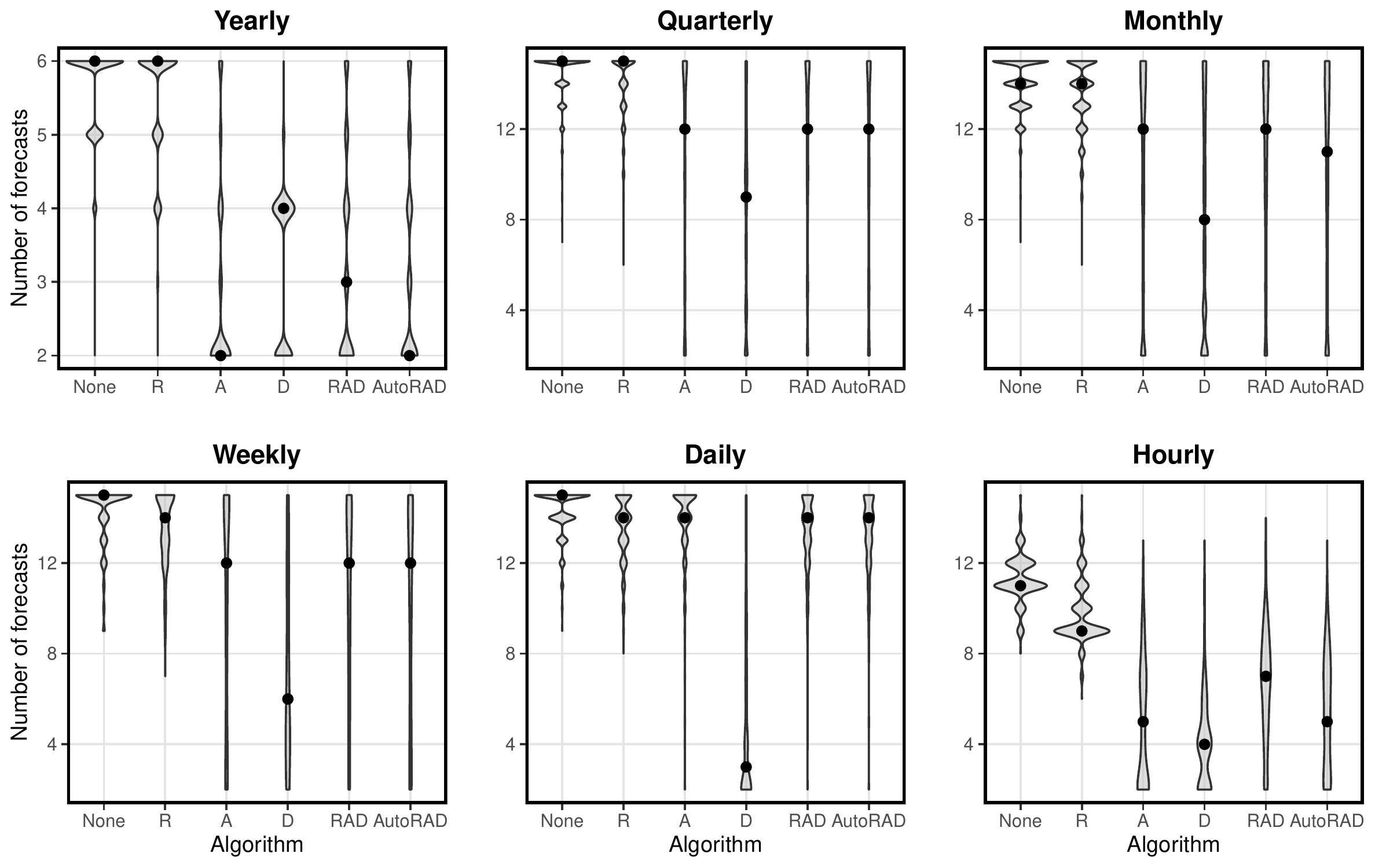}
  \caption{Number of individual forecasts retained in the optimal subsets identified by different trimming algorithms for each data frequency separately. The median values are depicted with black dots.}
  \label{fig:nof}
\end{figure}

To investigate the statistical significance of the performance differences, we perform the Multiple Comparisons with the Best \citep[MCB,][]{Koning2005-m3} test on each data frequency as well as across all frequencies. Note that we present the results after pooling time series from the forecasting competitions together. The performance differences between two algorithms are statistically significant if the intervals of the two algorithms do not overlap. The results of the MCB test based on MASE are presented in Figure~\ref{fig:mcb_mase}. The MCB results based on sMAPE were consistent with the ones based on MASE, and as such we do not present the sMAPE results in the paper for brevity. Overall, \method{RAD} and \method{AutoRAD} significantly outperform \method{None}, \method{R}, \method{A}, and \method{D} (see the top panel). The remaining six panels for different data frequencies show that \method{D} ranks worst in most data frequencies, apart from the daily data. Moreover, \method{RAD} or \method{AutoRAD} outperforms \method{None}, \method{R}, \method{A} and \method{D} in most cases. Therefore, \method{RAD} and \method{AutoRAD} can be recognized as guaranteed approaches for trimming an available forecast pool, and the simple average of the selected optimal subset poses a tough benchmark to beat. Additionally, as shown in Figure~\ref{fig:nof}, \method{RAD} and \method{AutoRAD} identify optimal subsets with relatively few individual forecasts from the original pool compared with other algorithms, thereby improving the computational efficiency of forecast combinations.

\subsection{Analysis}
\label{sec:analysis}

\subsubsection{The effect of the level parameter}
\label{sec:level}

\begin{figure}[!ht]
  \centering
  \includegraphics[width=\textwidth]{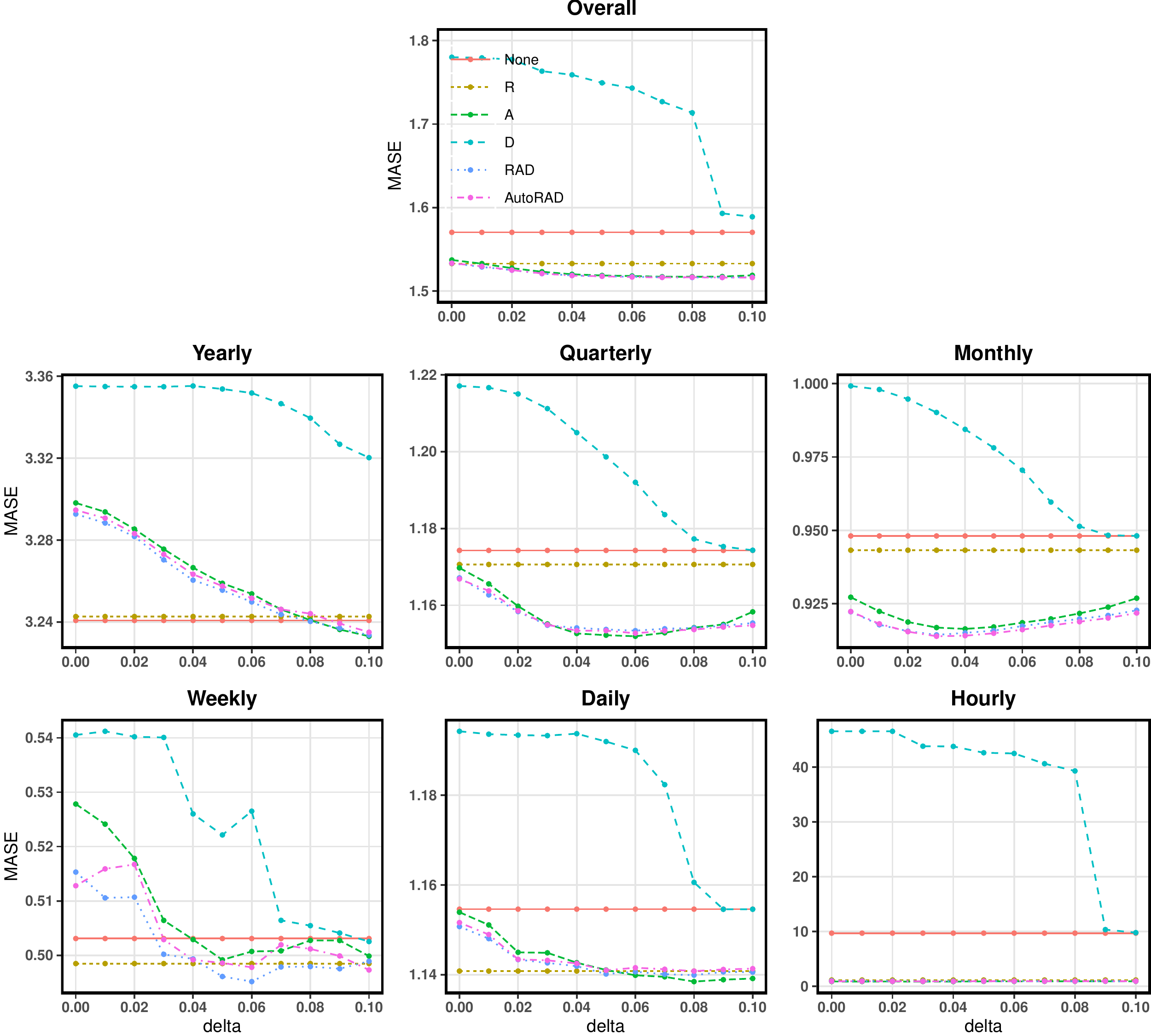}
  \caption{The effect of the level parameter $\delta$ on the performance of each forecast trimming algorithm for each data frequency separately and across all frequencies (Overall).}
  \label{fig:mase_delta}
\end{figure}

In Section~\ref{sec:RAD}, we introduced a level parameter, $\delta$, to identify whether the percentage drop in the ADT criterion is significant, which automatically determines the cut-off point at which we stop removing individual forecasts from the pool when using \method{RAD}. The level parameter is also used in the benchmark forecast trimming algorithms presented in Section~\ref{sec:benchmark}. The greater the $\delta$ value, the more difficult it is to eliminate an individual forecast from the forecast pool, that is, the number of individual forecasts in the selected optimal subset tends to be larger. To explore the importance of the level parameter $\delta$ when implementing a forecast trimming algorithm, we calculated the forecast error (in terms of MASE) of each trimming algorithm for various values of the level parameter, as depicted in Figure~\ref{fig:mase_delta}. Note that the performance of \method{None} and \method{R} does not vary with different values of $\delta$.

Overall, \method{RAD} and \method{AutoRAD} are superior to other four trimming algorithms across all values of $\delta$. A value of $\delta$ in the region between 0.04 and 0.06 seems to work well for seasonal series. Larger $\delta$ values would result in better performance for the yearly frequency. This may be caused by the small size of the original forecast pool (consisting of six models) for yearly series. When $\delta$ takes values between 0.04 and 0.06, trimming the forecast pool using \method{RAD}, \method{AutoRAD}, or \method{A} would yield better performance than using \method{None} or \method{D} for the quarterly, monthly, weekly, daily, and hourly frequencies. It is noteworthy that the average performance gap between \method{RAD} (or \method{AutoRAD}) and \method{A} is relatively small. \method{R} works well for the yearly, daily, weekly, and hourly frequencies, but performs poorly for the quarterly, monthly frequencies. Therefore, \method{RAD}, \method{AutoRAD} and \method{A} are more secure and reliable choices compared with other forecast trimming algorithms and a $\delta$ value between 0.04 and 0.06 is always recommended.

\subsubsection{Guidelines for selecting trimming algorithms}
\label{sec:guidelines}

As presented in Sections~\ref{sec:results} and \ref{sec:level}, \method{A} is a very competitive forecast trimming algorithm, which performs only slightly worse than the \method{RAD} and \method{AutoRAD} algorithms on average. In this section, we proceed by investigating the performance differences among \method{A}, \method{RAD} and \method{AutoRAD}, and then attempt to develop some empirical guidelines to determine which trimming algorithm to choose for a target time series.

To explore the importance of the degree of diversity relative to accuracy for a given pool on the selection of trimming algorithm, we propose an additional definition, denoted as RelDiv (Relative Diversity), which focuses on the ratio of diversity to accuracy for an available forecast pool. The RelDiv is given by
\begin{align}
  \mathrm{RelDiv} &= \frac{\mathrm{AvgMSEC}}{\mathrm{AvgMSE}} = \frac{\frac{1}{M^2}\sum_{i=1}^{M-1}\sum_{j=2, j>i}^{M}\mathrm{MSEC}_{i, j}}{\frac{1}{M}\sum_{i=1}^{M}\mathrm{MSE}_{i}} \nonumber \\
  &= \frac{\sum_{i=1}^{M-1}\sum_{j=2, j>i}^{M}\left[\frac{1}{H} \sum_{h=1}^{H}\left(f_{i, h}-f_{j, h}\right)^{2}\right]}{M\sum_{i=1}^{M}\left[\frac{1}{H} \sum_{h=1}^{H}\left(f_{i, h}-y_{h}\right)^{2}\right]}. \label{eq:reldiv}
\end{align}
The RelDiv measure is comparable between time series with different scales, and thus we can average the RelDiv values across time series.

We are interested in the percentage of cases in which \method{RAD} or \method{AutoRAD} outperforms \method{A}. We remove the instances in which both algorithms identify the same optimal subset from the given forecast pool. Then we split the time series with regard to different levels of RelDiv (low, moderate, and high levels) in Equation~\eqref{eq:reldiv} using the first and third quantiles of the sample values of RelDiv. In this study, the first and third quantiles are 0.23 and 0.53, respectively. The results from our analysis based on MASE are reported in Table~\ref{tab:pos}. Overall, the percentage of cases in which \method{RAD} or \method{AutoRAD} outperforms \method{A} is consistently greater than $50\%$ for the moderate and high RelDiv levels.

\begin{table}[!ht]
  \centering
  \caption{Percentages of series in which \method{RAD}, \method{AutoRAD}, \method{RAD} or \method{AutoRAD} outperform \method{A} for different levels of RelDiv (low, moderate, and high levels) in terms of MASE after excluding the instances in which both algorithms select the same optimal subset from the original forecast pool.}
  \scalebox{0.9}{
  \begin{tabular}{cccccccccc}
  \toprule
   & \multicolumn{3}{c}{\method{RAD}} & \multicolumn{3}{c}{\method{AutoRAD}} & \multicolumn{3}{c}{\method{RAD} or \method{AutoRAD}} \\
  \cmidrule(lr){2-4} \cmidrule(lr){5-7} \cmidrule(lr){8-10}
  Frequency & Low & Moderate & High & Low & Moderate & High & Low & Moderate & High \\
  \midrule
  Yearly & 50.1\% & 54.1\% & 49.7\% & 48.6\% & 55.7\% & 48.8\% & 49.0\% & 56.8\% & 49.9\% \\
  Quarterly & 50.0\% & 51.5\% & 47.6\% & 49.4\% & 52.1\% & 48.2\% & 49.9\% & 54.3\% & 51.0\% \\
  Monthly & 50.6\% & 53.4\% & 52.5\% & 50.2\% & 55.0\% & 54.9\% & 50.7\% & 56.9\% & 57.6\% \\
  Weekly & 52.7\% & 56.1\% & 58.7\% & 56.2\% & 55.0\% & 65.7\% & 56.2\% & 55.0\% & 65.7\% \\
  Daily & 38.4\% & 50.0\% & 57.7\% & 37.2\% & 55.3\% & 52.8\% & 37.2\% & 55.3\% & 55.6\% \\
  Hourly & 16.7\% & 44.7\% & 44.4\% & 25.0\% & 44.0\% & 45.5\% & 25.0\% & 48.0\% & 49.7\% \\
  \cmidrule(lr){2-10}
  Overall & 49.0\% & 53.0\% & 50.5\% & 48.0\% & 54.4\% & 51.5\% & 48.4\% & 56.2\% & 53.8\% \\
  \bottomrule
  \end{tabular}
  }
  \label{tab:pos}
\end{table}

\begin{figure}[!ht]
  \centering
  \includegraphics[width=\textwidth]{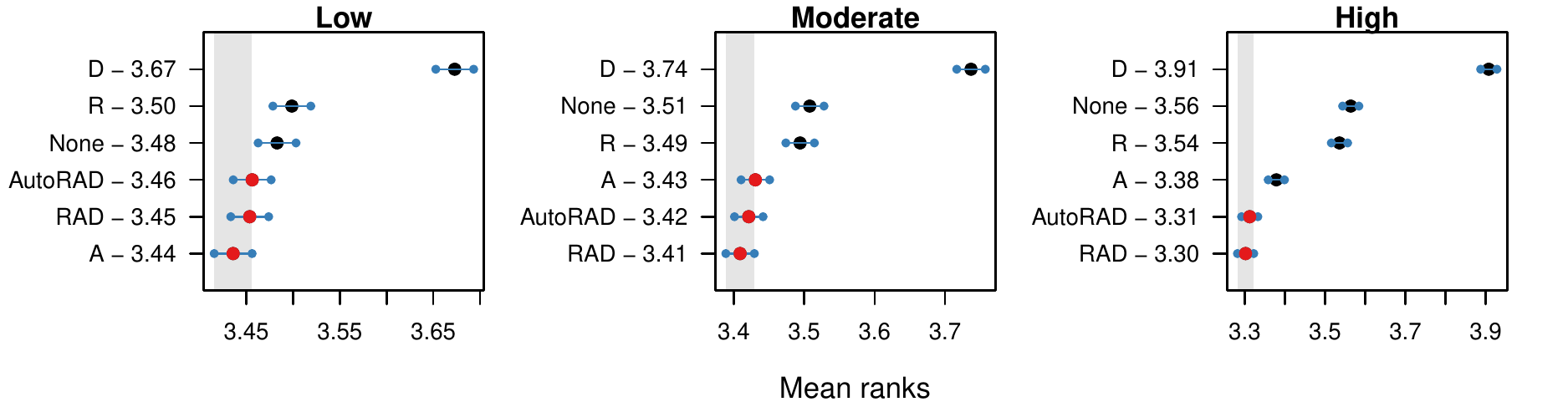}
  \caption{MCB tests on the ranks of the MASE values of the simple average forecasts remained after using \method{None}, \method{R}, \method{A}, \method{D}, \method{RAD} and \method{AutoRAD} for different levels of RelDiv.}
  \label{fig:mcb_ratio_mase}
\end{figure}

To investigate the statistical significance of the performance differences, we perform the MCB test on time series with low, moderate, and high RelDiv levels, respectively. Figure~\ref{fig:mcb_ratio_mase} depicts the results of the MCB test based on the MASE values. We observe that \method{A} is ranked first for the low RelDiv level, but its mean rank is not significantly different from \method{RAD} and \method{AutoRAD}. \method{RAD} achieves the best forecast accuracy for higher levels of RelDiv (moderate and high level). \method{RAD} and \method{AuoRAD} perform similarly for different levels of RelDiv, and the two algorithms have significantly better mean ranks than \method{A} for the high RelDiv level.

Given a time series and its forecast pool on both validation and test sets, the results of this study can serve as a basis to provide some guidelines to help forecasters select appropriate algorithms for forecast trimming. We list the guidelines below.
\begin{itemize}
  \item Although a considerable literature has accumulated over the recent years on discussing and emphasizing the importance of diversity, we do not necessarily have to address the diversity issue when selecting a forecast pool for combinations. There are specific conditions under which considering diversity is necessary and beneficial.
  \item For forecast pools with RelDiv values smaller than 0.2 on the validation set, \method{A} is the preferred option. Diversity does not need to be taken into account in forecast trimming algorithms in this case.
  \item \method{A}, \method{RAD} and \method{AutoRAD} behave similarly if RelDiv of the given pool takes values between 0.2 and 0.5. It is noteworthy that \method{A} and \method{RAD} will be more computationally efficient than \method{AutoRAD} as we do not have to determine the scale factor, $\kappa$, based on historical forecasting performance.
  \item \method{RAD} and \method{AutoRAD} are preferred if the RelDiv value of the given pool is greater than 0.5. \method{RAD} makes it a better choice due to its appealing computational efficiency.
\end{itemize}

\section{Trimming forecasts from different model families}
\label{sec:diffclass}

In the previous section, we investigated the performance of various forecast trimming algorithms when selecting individual forecasts from models within the exponential smoothing family. In this section, we proceed by applying these trimming algorithms to trimming a forecast pool consisting of models across different model families and investigate their performance.

The forecast pool we consider consists of nine individual forecasting methods, which are commonly used in recent forecast combination studies \citep[e.g.,][]{Montero-Manso2020-tq,Kang2020-rl,Wang2022-un}. Note that different settings are used for some methods, such as ARIMA, ETS, and NNET-AR. Prediction intervals of NNET-AR are computed using 1,000 simulations with normally distributed errors, which may be computationally expensive. The nine forecasting methods in our forecast pool are described in Table~\ref{tab:divmodels}, together with their \proglang{R} implementations.

\begin{table}[!ht]
  \centering
  \caption{The forecast pool consisting of methods across different model families.}
  \resizebox{\linewidth}{!}{
  \begin{tabular}{p{0.15\columnwidth}p{0.55\columnwidth}p{0.3\columnwidth}}
  \toprule
  Forecasting method & Description & \proglang{R} implementation \\
  \midrule
  NAIVE & The simplest time series forecasting method in which the point forecasts of all forecast horizons are equal to the last observation in the training period. & \code{forecast:naive()} \\
  SNAIVE & The seasonal na\"{i}ve method in which point forecast is equal to the most recent value of the same season. & \code{forecast:snaive()} \\
  RW-DRIFT & Random walk with drift. & \code{\specialcell[t]{forecast:rwf(..., \\ drift = TRUE)}} \\
  THETA & A decomposition approach to forecasting by modifying the local curvatures of the time series with Theta-coefficient \citep{Assimakopoulos2000-th}. & \code{forecast:thetaf()} \\
  ARIMA & The autoregressive integrated moving average model automatically selected by the AICc value \citep{Hyndman2008-au}. & \code{forecast::auto.arima()} \\
  ETS & The exponential smoothing state space model \citep{Hyndman2002-et}. & \specialcell[t]{\code{forecast::ets()} for $s \leqslant 24$ \\ \code{smooth::es()} for $s > 24$} \\
  TBATS & The exponential smoothing state space model with Trigonometric, Box-Cox transformation, ARMA errors, Trend and Seasonal components \citep{De2011-tb}. & \code{forecast::tbats()} \\
  STLM-AR & The seasonal and trend decomposition using Loess \citep{Cleveland1990-st} with an AR model fitted for the seasonally adjusted series. & \code{\specialcell[t]{forecast::stlm(..., \\ modelfunction = ar)}} \\
  NNET-AR & A feed-forward neural network using autoregressive inputs. & \code{forecast::nnetar()} \\
  \bottomrule
  \end{tabular}}
  \label{tab:divmodels}
\end{table}

We do not apply the pre-step described at the last paragraph of Section~\ref{sec:design} because the \pkg{forecast} and \pkg{smooth} \proglang{R} packages automatically search the ``best'' model based on specific information criterion and the output is relatively reliable. We present the results for each data set separately but also over all data sets in Table~\ref{tab:results_mcomp_div}. Similar to our results obtained from exponential smoothing, we observe that \method{RAD} and \method{AutoRAD} perform better than any of other trimming algorithms when we focus on the mean forecast errors (in terms of MASE, sMAPE and MSIS) across all data sets. The superiority is particularly evident when considering prediction intervals. \method{RAD} and \method{AutoRAD} achieve a balance between coverage and spread of prediction intervals. Moreover, the average performance gap between \method{A} and these two algorithms (\method{RAD} and \method{AutoRAD}) is small. Another important finding is that the overall performance of \method{None} based on different model families is worse than that of the exponential smoothing family, whereas the performance gap between the two different settings becomes smaller or even disappears when using \method{RAD} or \method{AutoRAD}. This finding confirms the guidelines documented in Section~\ref{sec:guidelines} that it is beneficial to address diversity as well as robustness and accuracy in forecast trimming algorithms (i.e., \method{RAD} and \method{AutoRAD}) when the available forecast pool has a high degree of diversity.

\begin{table}[!ht]
  \centering
  \caption{The average forecasting performance of each forecast trimming algorithm based on the pool of models from different classes for each data set and measure. The bottom panel also reports the overall results across all data sets. The best trimming algorithm is boldfaced for each data set and measure.}
  \scalebox{0.8}{
  \setlength\extrarowheight{-2pt}
  \begin{tabular}{ccp{0.125\columnwidth}p{0.125\columnwidth}p{0.125\columnwidth}p{0.125\columnwidth}p{0.125\columnwidth}p{0.125\columnwidth}}
  \toprule
   & & \multicolumn{6}{c}{\textbf{Simple Average}} \\
  \cmidrule(lr){3-8}
  Data set & Measure & \method{None} & \method{R} & \method{A} & \method{D} & \method{RAD} & \method{AutoRAD} \\
  \midrule
  M & MASE & 1.709 & 1.698 & \textbf{1.624} & 1.873 & 1.634 & 1.632 \\
   & sMAPE & 15.671 & 15.594 & \textbf{15.288} & 16.894 & 15.631 & 15.714 \\
   & MSIS & \textbf{20.898} & 20.923 & 21.865 & 24.060 & 21.943 & 21.987 \\
   & Coverage & \textbf{0.875} & 0.866 & 0.817 & 0.862 & 0.827 & 0.822 \\
   & Upper coverage & \textbf{0.905} & 0.902 & 0.892 & 0.894 & 0.899 & 0.896 \\
   & Spread & 0.950 & 0.912 & \textbf{0.794} & 1.076 & 0.818 & 0.810 \\
   & Bias & 0.087 & 0.084 & \textbf{0.058} & 0.102 & 0.061 & 0.059 \\
  \midrule
  M3 & MASE & 1.382 & \textbf{1.356} & 1.396 & 1.602 & 1.389 & 1.388 \\
  & sMAPE & 13.190 & \textbf{13.047} & 13.348 & 14.736 & 13.301 & 13.300 \\
  & MSIS & 13.103 & \textbf{13.069} & 15.053 & 15.752 & 14.602 & 14.747 \\
  & Coverage & \textbf{0.917} & 0.911 & 0.872 & 0.904 & 0.887 & 0.879 \\
  & Upper coverage & \textbf{0.939} & 0.936 & 0.925 & 0.929 & 0.931 & 0.929 \\
  & Spread & 0.876 & 0.857 & \textbf{0.790} & 0.992 & 0.823 & 0.808 \\
  & Bias & 0.022 & 0.021 & \textbf{0.002} & 0.037 & 0.003 & \textbf{0.002} \\
  \midrule
  M4 & MASE & 1.656 & 1.623 & \textbf{1.518} & 2.266 & \textbf{1.518} & \textbf{1.518} \\
  & sMAPE & 12.458 & 12.307 & 12.114 & 14.854 & \textbf{12.102} & 12.111 \\
  & MSIS & 16.626 & 16.507 & 16.081 & 23.097 & \textbf{15.790} & 15.969 \\
  & Coverage & \textbf{0.909} & 0.905 & 0.887 & 0.886 & 0.895 & 0.891 \\
  & Upper coverage & 0.931 & 0.929 & 0.934 & 0.911 & \textbf{0.937} & 0.935 \\
  & Spread & 1.121 & 1.103 & \textbf{0.858} & 1.757 & 0.876 & 0.870 \\
  & Bias & 0.051 & 0.048 & \textbf{0.021} & 0.084 & 0.022 & \textbf{0.021} \\
  \midrule
  Overall & MASE & 1.650 & 1.616 & \textbf{1.516} & 2.244 & \textbf{1.516} & \textbf{1.516} \\
   & sMAPE & 12.509 & 12.359 & 12.178 & 14.870 & \textbf{12.169} & 12.179 \\
   & MSIS & 16.571 & 16.456 & 16.109 & 22.906 & \textbf{15.817} & 15.993 \\
   & Coverage & \textbf{0.909} & 0.904 & 0.886 & 0.886 & 0.894 & 0.890 \\
   & Upper coverage & 0.931 & 0.929 & 0.933 & 0.912 & \textbf{0.936} & 0.935 \\
   & Spread & 1.113 & 1.095 & \textbf{0.856} & 1.729 & 0.874 & 0.868 \\
   & Bias & 0.051 & 0.048 & \textbf{0.021} & 0.083 & 0.022 & \textbf{0.021} \\
  \bottomrule
  \end{tabular}}
  \label{tab:results_mcomp_div}
\end{table}

\section{Conclusions and discussions}
\label{sec:conclusions}

In this paper, we proposed a simple and generic way to select an optimal subset from a forecast pool. Our approach is the first to take into account diversity for forecast trimming. Instead of focusing solely on diversity, we achieve a trade-off between accuracy and diversity of a given pool through a new criterion ADT. The proposed algorithm, \method{RAD}, addresses robustness, accuracy, and diversity simultaneously. In addition, we design another five forecast trimming algorithms as benchmarks to facilitate performance comparisons and future algorithmic development, including one trimming-free algorithm and several trimming algorithms isolating each one of the three key issues.

We showed empirically and analytically that the optimal subset identified using our approach achieves good performance and robustness in general in terms of both point forecasts and prediction intervals. Additionally, it offers balanced coverage versus interval spread that translates to a balance between achieved service level and holding costs.

Although a considerable literature has accumulated in recent years on discussing and emphasizing the importance of diversity, our analysis shows that we do not necessarily have to address diversity for forecast trimming. Instead, we offer some simple guidelines for selecting an appropriate forecast trimming algorithm for a given time series and its forecast pool. In brief, it is sufficient to focus only on accuracy in forecast trimming when the available pool has a low degree of relative diversity. However, in cases with a high degree of relative diversity, it would be a better to address robustness, accuracy, and diversity simultaneously.

The algorithms designed in our paper provide automatic tools for forecast trimming practice. They automatically determine the cut-off point at which we stop removing individual forecasts from the forecast pool. Moreover, these trimming algorithms require low computational cost because the calculations of the related criterion (e.g., ADT, AvgMSE, and AvgMSEC) are trivial. Using the forecast trimming algorithms, we identify an optimal subset from the original forecast pool, which helps to facilitate effective forecast combinations with improved computational efficiency.

In this paper, only univariate time series forecasting models are used to perform the performance comparison. The future work can consider to exam the forecast trimming algorithms using multivariate models. An equal-weighted combination is adopted in this research to evaluate the quality of the selected optimal subset identified using trimming algorithms due to its simplicity and robustness. Other weight estimation schemes can also be applied in the future but the choice of the scheme needs to be careful to make the performance of different trimming algorithms comparable. We focus on the forecast trimming for a single time series in our research. In other words, the proposed algorithms are series-specific. It is also possible to design several forecast trimming algorithms for a whole data set for the purpose of achieving effective forecast combinations through a global model in a cross-learning fashion.

\section*{Acknowledgments}

We thank the reviewers for helpful comments that improve the contents of this paper. This research was supported by the National Social Science Fund of China (22BTJ028).

\printbibliography

\end{document}